RESIDUE PROFILE IN PREDIVERGENCE SEQUENCES AS

A GUIDE TO THE ORIGIN OF DNA REPLICATION

(code evolution/replicase residues/DNA origin/
predivergence events)

Brian K. Davis

Research Foundation of Southern California, Inc.
5580 La Jolla Boulevard, PMB 60, La Jolla CA 92037 U.S.A.

2 Figures
21 Pages

ABSTRACT  DNA dependent RNA polymerase core subunits αββ' conserved residues at frequencies most closely matched with codons at stage 10.4-11.1 in code evolution. An excess of acidic residues lowered this estimate, but not by more than 2.3 stages. With 1 tryptophan (stage 14 addition) in 529 conserved residues,  αββ' significantly under-represented this amino acid, consistent with a cut-off in its residue profile before completion of the genetic code. Residue profiles in FEN-1 homologs and DNA topoisomerase I-5' placed their origin at stage 11-13. Prokaryote septation protein, FtsZ, arose earlier, between stage 8-11. Proteolipid in an ATP-driven proton pump served as a marker for cell membrane formation. It indicated this event took place near stage 7. Cell division, DNA replication and transcription were inferred to have originated in a protocell antecedent of the last common ancestor. Late-forming residue profiles characterized RNA dependent RNA replicase, DNA polymerase, reverse transcriptase and ribonucleotide reductase. This suggests some early processes, including RNA replication and deoxynucleotide synthesis, once depended on catalysts not found in extant residue sequences. Early formation of topoisomerase I, and enzymes that synthesize and trim RNA-DNA hybrids was viewed as evidence for a mixed duplex, with linear RNA and DNA strands, in the transition from an RNA to DNA genome.

Sequence homology in DNA dependent RNA polymerase and some accessory proteins, and the presence of a DNA genome in cells from each of the three domains of life suggest that DNA replication arose in the predivergence era.[1] As ribonucleotides produce deoxyribonucleotides, and RNA primes lagging strand synthesis during DNA replication and is integrated into translation and other cell processes, DNA synthesis evidently originated in a rioorganism, equipped with ribozymes and an RNA genome.[2-3] This scenario appears at variance, however, with the manifest heterogeneity of many replication proteins, including DNA polymerase.[5] To reconcile this with DNA replication in the last common ancestor, it is proposed that replication underwent substantial change, following separation of Bacteria from Archaea and Eukarya.[6] The prospect of gaining an insight into the time of origin of DNA replication, from a predivergence source, prompted a search for evidence of this transition in the residue profile of proteins formed before completion of the genetic code.

DNA synthesis preceding the 'standard code' could be expected to involve an ancestral replicase, deficient in late-comers to the code. The possibility of chronicling events in the pre-divergence era by means of this approach was recognized by Eck & Dayhoff.[7] Before the main stages of code evolution were known, they proposed that the simple residue profile in conserved segments of ferredoxin indicated this small protein had evolved when the code was incomplete. Residues in ancestral ferredoxin were recently shown to match the codon pattern at the fifth of 18 stages identified in code evolution.[8] In addition, conserved residues in the proteolipid subunit of a membrane proton pump ($H^+$-ATPase) were demonstrated to match codons at stage 7. This supported the proposition that the first cells, with a proteo-phospholipid membrane, formed about midway through code evolution. The origin of DNA replication is now examined. It will be seen that this process can be traced to a protocell antecedent of the last common ancestor.

In this investigation, proteins whose phylogenetic distribution establish that they formed before divergence from the last common ancestor are ordered in time, through comparing features in their residue profile with those expected at identified stages of code evolution. These stages were established after recognizing that the code had coevolved with amino acid biosynthetic pathways, which initially branched from the citrate cycle. With path lengths measured from this cycle, distinct types of amino acids were seen to enter the code at different stages[8]: $NH_4^+$ fixers (Asp, Glu, Asn, Gln), with paths of only 1 or 2 reactions, preceded small, increasingly hydrophobic molecules (Ala, Val, Pro, Ser, Gly, Cys, Thr, Leu, Ile, Met), with paths of 4 to 7 reactions, and they preceded residues bearing large basic or aromatic side chains (Arg, Lys, Phe, Tyr, His, Trp), whose paths extend for 9 to 14 reactions. A set of codon patterns resulted that clarified long-recognized regularities in the code.[9] They also matched the distribution of conserved residues in predivergence proteins.[8]

Two features of the residue profile were used to determine when an archaic sequence formed in the predivergence interval. First, the stage (path length) of its most advanced amino acid was noted. Where a sequence originated before completion of the code, its residue profile will exclude amino acids added to the code at a later stage. A sizeable difference in path length between the most advanced residue in a conserved sequence and Trp (stage 14) substantially decreases the probability that a sequence formed after completion of the genetic code,

$$\Phi = (1 - f)^N \, , \, f = \sum_{i=n+1}^{14} f_i \, , \, 0 \leq f \leq 1$$

$\Phi$ corresponds to the fraction of sequences containing $N$ residues, where none have a path length exceeding $n$ reactions. $f_i$ is the expected frequency of residues with path length $i$. Code evolution effectively advanced by 1 stage, with addition of 1 reaction to amino acid biosynthesis; each stage being estimated to require a notional interval of $10^6$ yr.[8]

A second indicator of protein 'age' relied on goodness of fit between residue frequencies and codon pattern at a given stage of code evolution. Agreement between them will tend to a maximum at the stage of origin, when selection independently targeted a sufficient number of residues. Present determinations involved a non-parametric measure, the Kolmogorov-Smirnov test. A single peak in goodness of fit, or an S-shaped curve resulted. They portray, respectively, sequence formation before and after code completion. Stage 14 effectively marked completion of the code. Subsequent changes were minor and produced comparatively small shifts in goodness of fit. With residue frequencies in 37 monomeric proteins[10] as reference, residues in predivergence sequences could be demonstrated to broadly approach the residue profile of a standard protein, as their time of origin advanced in relation to code evolution,

$$p = 0.0307 + 3.08 \times 10^{-2} S$$

where p is the probability that standard and sample distributions respresent the same underlying distribution and S signifies stage of origin; regression goodness of fit, $r^2$, was 0.14.

Some peak broadening accompanied advancement of the origin toward its upper detectable limit, at stage 14. An expansion of the sequence fixation interval would be anticipated, as the number of active-site residues increased in proteins of growing size and complexity. The tempo of change would also slow as variant formation decreased, with improved replication fidelity. A departure from 'one tRNA species per amino acid species', in the early code, would further obscure goodness of fit, following appearance of isoacceptors. Results from analysis of residue profiles were consistent with interspecies sequence homology. Residue profiles found to have evolved before completion of the genetic code occurred in proteins with wide phylogenetic distribution.

Figure 1 shows conserved residues in several proteins, distributed by their stage of entry to the genetic code. The proteins include prebiotic catalysts and DNA specific enzymes arrranged with an expanding residue range. Residues up to stage 5 occur in the protein ancestral to ferredoxin from the anaerobe, <u>Clostridium pasteurianum</u>,[7] and to stage 7 in the proteolipid subunit of $H^+$-ATPase (related to V-ATPase) from the crenarchaeote, <u>Sulfolobus acidcaldarius</u>.[11] It is extremely unlikely that a residue sequence with a stage 5 cut-off, as in ferredoxin, could form from the complete (stage 14) code[8] or a standard set of residue frequencies[10]; $6.55 \times 10^{-8} \leq p \leq 1.64 \times 10^{-6}$, the probability limits refer, respectively, to code and protein distributions. A stage 7 cut-off in the proteolipid residue sequence also departs significantly from the reference frequencies; $3.75 \times 10^{-2} \leq p \leq 6.50 \times 10^{-2}$. Other proteins in Fig. 1 conserve residues of later vintage, consistent with formation after ancestral ferredoxin and proteolipid.

Figure 1

Goodness of fit between conserved residue frequencies in the sequences examined and codon patterns, at each stage of code evolution, are given in Fig. 2. Consistent with the limited range of its residue profile, residue frequencies for ancestral ferredoxin optimally matched the codon pattern at stage 5.4 ± 0.31 (stage of best fit ($p_{max}$) ± standard error; S.E. = $\sigma/\sqrt{n}$, where $\sigma$ is the standard deviation (half-width of peak at $p_{max}$ x 0.606) and n the number of sample points forming the peak) Replacement of Val46 by Glu17, which coalign after superposition of segments from an early gene duplication,[7] gives a slightly more archaic sequence than that appraised.[8] Goodness of fit with stage 5 codons is then reduced, however, from 0.98 to 0.90. Also consistent with its residue range (Fig. 1), proteolipid residue frequencies matched most closely codons at stage 6.2 ± 0.45. Exclusion of charged and polar residues (stage 2 amino acids) from proteolipid (Fig. 1) conforms with strong selection forces for non-polar residues in forming this prebiotic membrane protein. When comparisons between proteolipid residue frequencies and codon patterns were restricted to stage 4-10, goodness of fit at stage 7 improved from 0.44[8] to 0.57 (Fig. 2a).

Figure 2

Proteolipid formation at stage 7 places it at a turning-point in code evolution. Following this event, the first positively charged and aromatic amino acids entered the code.[8] The antiquity of $H^+$-ATPase[12,13] and intramembrane location of proteolipid suggests that evolution of the first cells, encapsulated by a proteo-phospholipid membrane, altered the course of code evolution. Confinement of early molecular processes within a cell membrane would free them from the restriction to anionic reactants governing a primordial surface system.[8,14] Ancestral ferredoxin predated proteolipid, by present criteria, and a stage 5 residue profile (Fig. 1, 2a) ensured the protein was polyanionic. The dependence of precellular metabolism on charge attraction is further illustrated by the multianionic molecules in central metabolism.[14-16]

Cell division and replication proteins were found to originate after stage 7, as anticipated. Prokaryote septation protein, FtsZ, occurs in both Archaea and Bacteria [17] establishing it also occurred in their common ancestor. Stage 10-11 residues (Lys, Phe, Tyr) were the most advanced amino acids conserved by FtsZ(Fig. 1), although they were under-represented (9 of 194 conserved residues vs. 6 of 62 codons; p(binomial) = $2.36 \times 10^{-5}$). Omission of His+Trp, furthermore, indicates there was an early cut-off in the FtsZ residue profile; $2.68 \times 10^{-6} \leq p \leq 4.45 \times 10^{-4}$. Cell division may thus be traced back to at least stage 11 during code evolution. Evidence of cell division as early as stage $8.3 \pm 0.43$ is furnished by the correlation between FtsZ residue frequencies and codon patterns (Fig. 2a). These results fix the origin of the septation protein at stage 8-11, possibly soon after appearance of the first cells.

A FEN-1 homolog, DNase IV, and its homologous domain in DNA pol I provided the first evidence of DNA synthesis. This enzyme targets the RNA-DNA hybrid molecule formed during lagging strand synthesis in DNA replication. A structural and functional homology between the 5'-exonuclease of Bacteria and Eukarya[18] places this enzyme in their common ancestor. FEN-1 homologs conserved residues up to stage 11 (Fig. 1). The probability of this cut-off is, $4.18 \times 10^{-2} \leq p \leq 1.48 \times 10^{-1}$. Its conserved residue frequencies were most compatible with codons at stage $8.7 \pm 1.05$ (Fig. 2a). These results indicate ancestral FEN-1 originated between stage 9-11.

Selection for acidic residues noticeably skewed the residue profile of this nuclease toward an earlier codon pattern. Structural studies on a phage FEN-1 homolog revealed that Asp clusters bind divalent cations (Zn, Mn) at the catalytic site, in front of a helical arch. Several Asp residues also occur in a motif behind the helical arch, where they would facilitate product release.[19] Stage 2 residues (Asp, Glu, Asn, Gln) form the largest group of amino acids conserved by FEN-1 homologs (Fig. 1). They account for 1/4 of all conserved residues (12/48), with acidic residues representing 3/4 of them. This exceeds the frequency of stage 2 codons in a stage 9-11 code, where they contribute 1/5 to 1/6 of assigned codons.[8] Stage 2-free goodness of fit determinations between conserved residues and assigned codons shifted $p_{max}$ to 11.5 ± 0.64. This advances the estimated time of origin for ancestral FEN-1 by 2.8 stages, placing it at the upper limit of the previous range and near the cut-off in nuclease residue range.

Consistent with these results, non-conserved residues in the 5'-exonuclease domain of DNA pol I[18] included residues to stage 13; 2 stages later than for conserved residues. Frequencies among 167 non-conserved residues displayed maximum goodness of fit for the codon pattern at stage 11.4 ± 0.75; 2.7 stages later than for conserved residues. Both parameters reveal that non-conserved sites in the nuclease sequence evolved later than its conserved, active sites.

Conserved sequences in topoisomerase and core subunits of DNA dependent RNA polymerase provide further evidence of early DNA replication. Topoisomerase I-5' from 8 bacteria species, including thermophiles Thermotoga maritima and Fervidobacterium islandicum, and the thermophilic euryarchaeon, Methanococcus jannaschii, contained residues up to stage 11 (Fig. 1) in motifs 1, 7a,b and Zn binding motifs 1-3.[20] This cut-off has a probability, $6.22 \times 10^{-2} \leq p \leq 1.88 \times 10^{-1}$. Residue frequencies correlated optimally with codons at stage 12.1 ± 0.75 (Fig. 2a). Topoisomerase I-5' in reverse gyrase (carboxy domain) of Sulfolobus shibatae B12 and topA gene product of T. maritima,[21] by comparison, conserved residues to stage 13, in motifs 1-10, with frequencies that best fit codons at stage 12.5 ± 0.84. These results fix the probable origin of topoisomerase I-5' between stage 11-13.

Topoisomerase I-5' is present in both prokaryote domains,[21] while topoisomerse II occurs in all 3 domains.[22] Both topoisomerases clearly originated before divergence of the prokaryote domains. Residue sequences in topoisomerase II protomers, ParC and ParE,[23] indicate, however, that they formed after the standard code (Fig. 2a). As topoisomerase I preceded decatanase, according to these results, genes apparently consolidated into linear chromosomes, which later became circular in the last common ancestor.

Each core subunit of DNA dependent RNA polymerase exhibits a goodness of fit peak in Fig. 2b. Subunit D ($\alpha$ analog, subunit binding domain) of archaeons S. acidocaldarius and Haloarcula marismortui and its eukaryote homologs from Saccharomyces cerevisiae, subunits AC40, B44,[24] display optimal agreement with codons at stage 9.6 ± 0.75. His is its most advanced residue (Fig. 1). A comparable profile arises in $\alpha$ of Escherichia coli and B. subtilis polymerases,[25] albeit $\beta\beta'$ binding domain residues predated those for the UP-promoter. Among 314 residues, 144 were conserved (identity, 0.46). Its most advanced residue being His (stage 13). $\alpha$ residue frequencies best fit the codon pattern at stage 10.4 ± 0.92. The $\beta'$ subunit of E. coli and its eukaryote analogue in S. cerevisiae, rpo 21 gene product,[26] conserved residues that matched codons at stage 11.1 ± 0.85 (Fig. 2b). A single Trp occurred among 142 conserved residues. ß subunit from E. coli, rpoB product, and S. cerevisiae rpb2 product,[27] likewise, conserved residues that best fit codons at stage 11.0 ± 0.78 (Fig. 2b). Conserved residues in β extended to His (Fig. 1).

RNA polymerase subunits αββ' are seen to conserve residues with a stage 10-11 frequency distribution. Moreover, the single Trp in αββ' (1/529 conserved residues) significantly under-represented this amino acid; $5.16 \times 10^{-7} \leq p \leq 1.67 \times 10^{-3}$. No Trp occurred among conserved residues (0/165) in archaeon RNA polymerase subunit D. These findings show RNA polymerase originated in a protocell, with a stage 10-13 code.

Sequence homology in RNA polymerase of Archaea, Bacteria and Eukarya[28] established early that transcription occurred in the last common ancestor.[29] Its polymerase residue profile now furnishes a probable time of origin for transcription within the prediveregence era. As with FEN 1 homologs, an excess of stage 2 residues skewed estimates based on the polymerase residue profile toward an earlier origin. Goodness of fit determinations, with stage 2-free distributions, advanced $p_{max}$ for each subunit by up to 2.3 stages. Broadly, this placed the origin of RNA polymerase at stage 11-13. This agrees with the profile cut-off. It also supports the finding that transcription evolved when the genetic code was still incomplete.

These stage 2-free goodness of fit determinations indicated that ß' was formed first among RNA polymerase subunits ($p_{max}$ = 11.7 ± 0.41). Mutational analysis[30] formerly revealed that ß' contains binding sites for both DNA template and RNA transcript growth point (3'-terminus). Sequence homologies,[31] furthermore, show a likely kinship between β' and reverse transcriptase (simian AIDS virus), τ subunit of DNA polymerase, and a number of other proteins that interact with DNA, or RNA.

DNA pol I,[32] plus-strand virus RNA dependent RNA replicase and prokaryote reverse transcriptase[33] have S-shaped goodness of fit curves (Fig. 2b), consistent with formation after the standard code. Profile cut-offs observed in these proteins were, respectively, stage 13, 11 and 14; His+Trp omission from RNA replicase was not significant, 0.18 ≤ p ≤ 0.36. A late-forming residue profile, with no visible goodness of fit peak (Fig. 2a), also occurred in type I and II ribonuclease reductase[34] and, as noted, in topoisomerase II protomers ParC and ParE (Fig. 2a). They conserved residues to stage 13-14. The glycyl radical domain (type III) of an <u>E. coli</u> type II ribonuclease reductase, with an adenosylcobalamin cofactor,[34] displayed a late-forming frequency distribution. It conserved residues up to stage 11; this cut-off had probability, 0.09 ≤ p ≤ 0.24. Its oxygen sensitivity, and S-adenosylmethionine and FeS cluster cofactors favour type III ribonucleotide reductase being ancestral to type I and II reductases, whose allosteric control mechanism it shares.[35] Present results support the view[29,35] that extant reductases were preceded by an earlier catalyst. It putatively catalyzed deoxynucleotide synthesis before completion of the genetic code. Late-forming sequences also characterized the 3'-5' nuclease domain of DNA polymerase - Klenow fragment,[32] clamp loading factor RF-C - DNA pol τ,[36] RNase H1,[37] and the ribo-proteozymes telomerase[38] and RNase P.[39]

This analysis of residue profiles in archaic proteins has provided evidence of a predivergence mechanism of DNA replication. Significant gaps remain, however. Some late-forming replication proteins almost certainly took over functions once performed by other molecules, not found in the record preserved by extant residue sequences. RNA apparently provided the infrastructure for replication and translation in pre-code and pre-cellular systems,[2-4,8,16,40] yet RNA replicase formed later than the standard code. Its formation was presumably retarded by an early reliance on a ribozymal replicase. Deoxynucleotide synthesis by a free radical reduction mechanism appears beyond the scope of a ribozyme.[35] However, a proto-reductase involving glycyl and Cys-thiyl free radicals, with adenosyl and FeS-cluster cofactors, could, in principle, have formed as early as stage 5 in code evolution. Formation of a nuclease specific for RNA-DNA hybrids and a DNA dependent RNA polymerase, by stage 11-13, conforms with early occurrence of a mixed polymer of DNA and RNA strands. Since reverse transcriptase evidently formed after the standard code (Fig. 2b), RNA polymerase specificity may have been modified[41] to catalyze RNA dependent DNA synthesis. Early formation of topoisomerase I favours a linear molecule as the hybrid intermediate in the transition from an RNA to DNA genome.

Figure legends

Fig. 1. Distribution of residues conserved by archaic proteins. The range of residues, with respect to the stage of addition to the code, is an indicator of when a protein formed in the predivergence era. Residue profiles are shown for proteins catalyzing DNA replication, monomer synthesis and transcription, and prokaryote cell division. Proteolipid subunit of a membrane proton pump from an acidothermophilic archaeon and ferredoxin from an anaerobic eubacterium exemplify very early, prebiotic proteins. At top, amino acids in groups, whose codons were assigned at the same stage of code evolution. Transition between each stage effectively corresponded to addition of one reaction step to an amino acid biosynthetic pathway; path length being measured from the citrate cycle. Each stage of code evolution was estimated to represent a notional interval of $10^6$ yr.[8] Frequency of conserved residues in sequences examined appear under Fig. 2.

Fig. 2. Probability curves peak at the stage of code evolution with best fit between assigned codon pattern and frequency distribution of conserved residues in each archaic protein. (a) Prebiotic, cell division, deoxyribonucleotide synthesis and accessory proteins, and (b) enzymes in DNA or RNA replication, DNA transcription and reverse transcription. DNAP, DNA polymerase (pol I-like) catalytic

site of Klenow fragment,[32] n = 138, π = 0.67, where n is number of sites compared, and π the fraction of sites with identical residues. Fdn, ferredoxin of C. pasterianum,[7] n = 25, π = 0.48. FEN-1, 5'-exonuclease of human and E. coli,[18] n = 215, π = 0.22. FtsZ, filamentous temperature sensitive mutant protein of 8 bacteria and 1 archaeon species and 1 chloroplast source,[17] n = 390, π = 0.49. ParC and ParE, partition proteins C and E of Mycoplasma homini and E. coli,[23] n = 717, π = 0.30 (ParC), n = 602, π = 0.37 (ParE). PL, proteolipid subunit of $H^+$-ATPase of S. acidocaldarius,[11] n = 11, π = 0.73. RNAP, DNA dependent RNA polymerase subunits β, β' and D: β of E. coli, rpoB product, and yeast, rpb2 product,[27] n = 342, π = 0.42; β' of E. coli and yeast, rpo21 product,[26] n = 324, π = 0.44; D of 2 archaeon species and yeast subunits AC40, B44,[24] n = 265, π = 0.62. RNAR, RNA dependent RNA replicase of 15 plus strand viruses,[33] n = 168, π = 0.16. RNR, ribonucleotide reductase of E. coli (class I) and T. acidophila (class II)[34] n = 366, π = 0.24; RT, reverse transcriptase of Myxococcus xanthus (msMx162, msMx65) and E. coli (msEc67, msEcB86),[33] n = 178, π = 0.83. Topo I, type I-5' topoisomerase of S. shibatae B12 and T. maritima,[21] n = 230, π = 0.50. p, probability that residue and codon frequency distributions were from same underlying distribution, with goodness of fit determined by a Kolmogorov-Smirnov test.